# Metasurfaces Leveraging Solar Energy for Icephobicity

*Efstratios Mitridis, Thomas M. Schutzius\*, Alba Sicher, Claudio U. Hail, Hadi Eghlidi\*, and Dimos Poulikakos\**

Laboratory of Thermodynamics in Emerging Technologies, Department of Mechanical and Process Engineering, ETH Zurich, Sonneggstrasse 3, CH-8092 Zurich, Switzerland.

**Abstract**

Inhibiting ice accumulation on surfaces is an energy-intensive task and is of significant importance in nature and technology where it has found applications in windshields, automobiles, aviation, renewable energy generation, and infrastructure. Existing methods rely on on-site electrical heat generation, chemicals, or mechanical removal, with drawbacks ranging from financial costs to disruptive technical interventions and environmental incompatibility. Here we focus on applications where surface transparency is desirable and propose metasurfaces with embedded plasmonically enhanced light absorption heating, using ultra-thin hybrid metal–dielectric coatings, as a passive, viable approach for de-icing and anti-icing, in which the sole heat source is renewable solar energy. The balancing of transparency and absorption is achieved with rationally nano-engineered coatings consisting of gold nanoparticle inclusions in a dielectric (titanium dioxide), concentrating broadband absorbed solar energy into a small volume. This causes a >10 °C temperature increase with respect to ambient at the air–solid interface, where ice is most likely to form, delaying freezing, reducing ice adhesion, when it occurs, to negligible levels (de-icing) and inhibiting frost formation (anti-icing). Our results illustrate an effective unexplored pathway towards environmentally compatible, solar-energy-driven icephobicity, enabled by respectively tailored plasmonic metasurfaces, with the ability to design the balance of transparency and light absorption.

Icing is very common in nature and technology, and when not controlled or alleviated, it can have very negative consequences in a broad range of applications including automobiles,[1] aviation,[2] power distribution,[3] shipping,[4] road transportation networks,[5] buildings,[6] wind energy generators[7] and photovoltaics.[8] The energy requirements, cost and environmental impact of de-icing are surprisingly high. A very common example from everyday life is automobile windshield defrosting by means of hot air, that mandates the function of the engine for up to 30 min in cold climates.[9] It is also estimated that the global aircraft de-icing market will be worth $1.30 billion by 2020,[10] while the ice protection systems are expected to amount for $10.17 billion by 2021.[11] A degree of optical transparency is a crucial property in many of these applications,[1,2,6,8] achieved by –often– multifunctional windshields and windows, constituting indispensable architectural elements of commercial and residential buildings.[12,13] Several passive anti-icing strategies, based on scientifically nano-engineered surfaces,[14,15] have been developed since the late 1990s, including hierarchical superhydrophobic surfaces,[16–22] lubricant infused surfaces,[23] and liquid infused polymers.[24] Their icephobic properties are defined by the nucleation delay, droplet contact time reduction, reduced ice adhesion and defrosting time.[16–24] While promising and highly desirable, passive approaches are to date complementary to active systems such as resistive heaters[7] and mechanical scraping.[4] Such active systems, however, are energy intensive (requiring electricity),[25] their operation is intrusive, have limited optical transparency and working temperature,[15,26] or require replenishment.[15,23] What is less explored is harvesting the potential of ubiquitous sunlight to impart icephobicity, here through specifically tailored plasmonic metasurfaces.

We show that by using rationally nano-engineered ultra-thin hybrid plasmonic metasurfaces, one can concentrate naturally occurring solar energy into a small volume, causing a greater than 10 °C temperature increase with respect to ambient at the air–solid interface, where ice is most likely to form, delaying freezing, reducing ice adhesion to negligible levels (de-icing) and inhibiting frost formation (anti-icing). Significant thermal responses can be achieved with transparent metasurfaces, paving the way to a wide range of applications where the benefit of icing resistance must be weighed against loss of transparency. We realize this by engineering an array of nanoscale noble metal particles embedded in a dielectric matrix –an approach that is shown to be capable of tuning absorption and transparency in a systematic way based on film thickness and is well-suited for a fundamental study– on industrially and commercially relevant substrates (*e.g.*, glass, plastic), for a total film thickness in the sub-micron regime, ensuring maximum temperature boost. The use of plasmonics in metal–dielectric composites in applications such as water desalination,[27] photovoltaics,[28] solar water heating[29] and photochemistry[28,30] were explored before. Plasmon resonances in metallic nanostructures can be damped radiatively (photon re-emission) or non-radiatively *via* Landau damping,[31] resulting in rapid localized heating of the nanoparticles.[32–34] Here, exploiting the Landau damping of hot electrons[32] in deeply sub-wavelength gold particles, and incorporating them in rationally designed metal-dielectric nanocomposite metasurfaces, we show that a broadband absorption of solar energy, with adjustable levels of absorption and transparency, can be achieved within sub-wavelength films, confined in the surface. We demonstrate that such heating can be considerable and under harsh icing conditions it can significantly delay frost formation (anti-icing) and lead to the swift removal of frozen ice blocks from the surface (de-icing) for several freezing cycles. This is a straightforward approach that leverages naturally occurring sunlight to achieve an impressive anti-icing and de-icing performance that does not rely on chemicals, mechanical action or electricity, translating into environmental and cost savings and operational facility.

**Results and Discussion**

We designed and fabricated plasmonic metasurfaces consisting of closely-packed, deeply sub-wavelength metal particles, in a dielectric matrix.[35,36] For the metal and dielectric, we chose gold (Au) and titanium dioxide (TiO$_2$), respectively. Nanoscopic gold particles are very effective absorbers of sunlight at their plasmon resonance wavelengths. Embedding them in a dielectric matrix, with a volumetric concentration close to the percolation limit, leads to a very effective and ultra-broadband absorption. This absorption is attributed to the significantly increased imaginary part of the effective permittivity, Im($\varepsilon_{r,nc}$) –equivalently effective electronic conductivity– over an ultra-broad spectrum,[35] in our designed nanocomposites. The increased conductivity leads to a boost in the photoexcited hot carriers,[32] which generate heat through Landau damping.[31] The selection of TiO$_2$ as the dielectric is based on previous findings in which high absorption levels of over 80% across the visible wavelength range were demonstrated in Au–TiO$_2$ thin films. The TiO$_2$ enhances the plasmon resonance of individual Au nanostructures and the plasmonic coupling of proximal nanostructures, enabling broadband light absorption.[37] Other common dielectric materials, such as silicon dioxide or Teflon, are also employable, albeit using the high-refractive index TiO$_2$ enables enhancement of the visible light absorption (see Supporting Information, section 'Modeling light absorption' and Figure S1 for a comparison between TiO$_2$ and other common dielectrics).

Based on our theoretical evaluations and experimental results (discussed below), a metasurface composed of gold particles with sizes, $d$, of $\approx 5$ nm, and a volumetric concentration, $v_{Au}$, of $\approx 40\%$, embedded in a TiO$_2$ matrix, exhibits a high level of absorption across the entire visible and near-infrared spectrum, and is the selected material system in this study.

To realize the nanocomposites, we deposited Au and TiO$_2$ *via* a layer-by-layer sputter deposition process, on fused silica and acrylic (PMMA) substrates. An experimental parametric study by changing the total film thickness ($L$) and the total number of deposited layers ($N_L$) was conducted for achieving a gold volumetric concentration, $v_{Au} \approx 40\%$, and a desired level of absorption and transparency. The individual layer thicknesses of Au and TiO$_2$ were kept constant at 4.6 and 6.9 nm, respectively. Due to the anti-wetting properties of Au, the thin deposited gold layers formed isolated particles instead of a continuous film, effectively leading to a nanocomposite with sub-wavelength, highly-absorptive inclusions (see Supporting Information, section 'Modeling light absorption' for the effect of the particle size). $N_L$ was varied from 4 to 44, producing film thicknesses ranging from $L = 38 \pm 2$ nm to $L = 270 \pm 5$ nm, respectively, including a 15-nm TiO$_2$ top-layer that provided identical surface chemistry for all the fabricated samples. To enhance film–substrate adhesion, a 2-nm chrome layer was deposited between the substrate and the metasurfaces.

Figure 1a shows a picture of a plasmonic metasurface fabricated by the above procedure ($L = 270$ nm). It also details representative cross-sectional and top-view scanning electron micrographs of the coating. The cross-section consists of layers of Au nanoparticles embedded in a TiO$_2$ matrix (shown with arrows). The pitch between adjacent Au layers, $p$, is constant ($p = 11 \pm 1$ nm). To analyze the gold particle size distribution, we acquired a top-view scanning electron micrograph of an Au layer deposited on a TiO$_2$ layer (Figure 1a, bottom-right; 8-layer metasurface, $L = 45$ nm, without a TiO$_2$ top-layer). Figure 1b shows a histogram of the number of Au nanoparticles, $N$, normalized by the total number of particles, $N_0$, vs. their equivalent diameter, $d$, assuming spherical particles. We found that the particle size has a gamma distribution with a mean value of 5.4 nm, variance of 2.1 nm and a range of 7.9 nm. The methodology of the particle size analysis is described in Supporting Information, section 'Nanoparticle size analysis'; see also Figure S2. In Figure 1c, the transparency of the metasurface (here, $L = 60$ nm) relative to a control sample is demonstrated. The metasurface was placed on top of a printed logo and illuminated with white light on the back side. Figure 1d shows the normalized light absorption spectra, $\mathcal{A}$, vs. wavelength of light, $\lambda$, for four metasurfaces with $L = 38$ nm, 60 nm, 95 nm, and 270 nm (wavelength range of 400–800 nm). The mean absorption can be calculated as:

$$\overline{\mathcal{A}} = \left[ \int_{\lambda_{min}}^{\lambda_{max}} \mathcal{A}(\lambda) d\lambda \right] / (\lambda_{max} - \lambda_{min}),$$ where $\lambda_{min} = 400$ nm, and $\lambda_{max} = 800$ nm. From Figure 1d we can see that the metasurfaces absorb light broadly across the visible spectrum and have a mean absorption value of $\overline{\mathcal{A}} = 28\%$, 37%, 63%, and 83% for $L =$ 38 nm, 60 nm, 95 nm, and 270 nm, respectively, indicating the tunability of $\overline{\mathcal{A}}$ with $L$. The broadband absorption is a result of the small sizes of the nanoparticles ($d << \lambda$) and their collective behavior ($p$ and $d$ are comparable) in the nanocomposite with a volumetric concentration close to the percolation limit. Figure 1e shows a plot of the normalized transmission spectra, $\mathcal{T}$, vs. $\lambda$, for the same four nanocomposites used in Figure 1d. Here we see that for $L =$ 38 nm, 60 nm, 95 nm, and 270 nm, $\overline{\mathcal{T}} = 51\%$, 36%, 15%, and 2%, designating also the tunability of $\overline{\mathcal{T}}$ with $L$. Information regarding the individual reflection and transmission spectra can be found in Supporting Information, section 'Optical spectroscopy' and Figure S3. Figure 1f shows the spectra of the absorbed, $\mathcal{A} \cdot I_r$, and transmitted, $\mathcal{T} \cdot I_r$, sunlight (standard solar irradiance at sea level) vs. $\lambda$, for the partially transparent metasurface with $L = 60$ nm. Also shown is the reference standard solar irradiance, $I_r$ ($\mathcal{A} = 0\%$ and $\mathcal{T} = 100\%$). This nano-engineered metasurface exhibits a good balance of transparency and absorption, all within a deeply-subwavelength film.

Figure 2a shows a schematic of the experimental setup we used to characterize the temperature change of the nano-engineered metasurfaces (deposited on glass substrates) due to visible light illumination. For illumination we used a halogen light source, which we collimated and focused on the metasurfaces. (See Supporting Information, section 'Thermography' and Figure

S4 for further information on characterizing plasmonically enhanced light absorption heating.) The temperature increase in the metasurface relative to ambient, $\Delta T$, was measured with a high-speed infrared camera (spectral range of 1.5 to 5.1 µm). The focused light diameter, $D$, was $6.0 \pm 0.3$ mm and the power density, $P$, was $2.4 \pm 0.2$ suns (kW m$^{-2}$). We used a mechanical shutter to rapidly control illumination. Figure 2a (inset) also shows the spectrum of the broadband light source used in this study. Figure 2b shows $\Delta T$ vs. time, $t$, for the four samples with different values of $\overline{\mathcal{A}}$ (28%, 37%, 63%, and 83%). Time-zero was when the metasurface was first illuminated. Here, $\Delta T$ was measured at the center of the illuminated area on the surface. It is clear that all metasurfaces exhibit an appreciable change in temperature due to visible light illumination, valid even for the highly transparent metasurfaces. We also note that there are transient and steady-state regimes for $\Delta T$. $\Delta T$ vs. $t$ curve is also shown for the reference sample (uncoated glass substrate). Figure 2c shows the corresponding spatial distributions of $\Delta T$ for the four metasurfaces at $t = 180$ s (steady state). The boundary of the illuminated area is marked with a dashed circle, and it is evident that the maximum value of $\Delta T$ occurs at the center of this area and that heat diffuses well beyond the illuminated area.

The time to steady state is controlled by the characteristic length, $L_\text{C} = (L_2 - D)/2$, and the substrate thermal diffusivity, $\alpha$, where $L_2$ is the side-length of the square sample. This time can be estimated as $L_\text{C}^2 / \alpha$. Substituting appropriate values yields $(6 \text{ mm})^2 / 0.43 \text{ mm}^2 \text{ s}^{-1} \approx 84$ s, which is comparable to the order of magnitude of the experimentally determined time ($\approx 100$ s). The temperature increase at the surface of the sample is determined by light irradiance and absorption as well as heat losses due to conduction (in the film and substrate), convection (in the surrounding air), and radiation (to the surrounding environment); see Supporting Information, section 'Heat Transfer' and Figure S5 for a detailed analysis on the above. In summary, to under-stand the relative importance of convection and radiation on determining the steady-state value of $\Delta T$, we solved for the temperature distribution (sample with $\overline{\mathcal{A}} = 37\%$) in a two-dimensional semi-infinite plate immersed in a gas that had a heated gas–substrate interface and an adiabatic condition on its bottom interface. The boundary condition at the interface was modified to account for radiation losses. We fixed a value of emissivity ($\varepsilon \approx 0.8$), based on the infrared measurements of $\Delta T$, and varied the position of $T_\infty$ (aspect ratio of $L_1/L_2$) until $T_s - T_\infty$ (the temperature difference between the gas–substrate interface and the gas very far away) matched our experimentally determined value of $\Delta T$. Based on the steady-state value of $\Delta T$ that we measured, we have determined that the percentage of cooling due to radiation is a mediocre 4% of the amount of heat provided by illumination, but it can also exceed 25% in certain cases and metasurfaces (see Supporting Information, section 'Heat Transfer'). To understand the importance of natural convection on cooling, we computed the Rayleigh number, $Ra_{L_1} = g\beta\Delta T L_1^3/(\nu\alpha)$, where $g$ is the acceleration due to gravity, $\beta$ is the expansion coefficient ($1/T$ for an ideal gas), and $\nu$ is the kinematic gas viscosity. Below and above a critical value of $Ra_{L_1}$, $Ra_c = 1708$,[38] heat transfer is in the form of conduction and advection (convection and conduction), respectively. Substituting appropriate values, we see that the value of $Ra_{L_1}$ in this case is $7\cdot10^{-3}$ ($Ra_{L_1} \approx 7\cdot10^{-3}$ for $T_\infty = 23$ °C, $P \approx 2.4$ kW m$^{-2}$, $\overline{\mathcal{A}} = 37\%$, $\Delta T \approx 7$ °C, $L_1/L_2 \approx 1.2\cdot10^{-2}$, $\nu = 1.57\cdot10^{-5}$ m$^2$ s$^{-1}$, $\alpha = 22.07\cdot10^{-6}$ m$^2$ s$^{-1}$, $L_1 = 212$ μm, $\beta = 3.38\cdot10^{-3}$ K$^{-1}$, $g = 9.81$ m s$^{-2}$).[39] Therefore, due to the fact that $Ra_{L_1} < Ra_c$, we conclude that heat transfer through conduction is the dominant mechanism and natural convection can be neglected.

Figure 3a shows a schematic of the experimental setup used to investigate the effect of illumination (halogen lamp; $D = 6.0\pm0.3$ mm, $P = 2.4\pm0.2$ kW m$^{-2}$) on film–ice adhesion (de-icing), on each metasurface, which was held in place by a holder on an x-y piezo stage. A hydrophobic cylinder (inner radius of $R = 1.5$ mm), filled with water, was placed on top of it, concentrically to the illuminated area. The experiments took place at $T = -4$ °C: a pin connected to a piezoelectric force sensor was initially pressed at the base of the frozen ice cylinder parallel to the x-axis inducing a shear stress, $\tau_{yx}$. (The value of $\tau_{yx}$, $\tau_{yx} = 90 \pm 5$ kPa, was selected to be close to, but less than, the mean ice adhesion strength of a PVDF-coated substrate, $131 \pm 19$ kPa; 3 experiments), in order to prevent premature detachment of the ice cylinder. Illumination was switched on with a mechanical shutter at $t = 0$ and $\tau_{yx}$ vs. $t$ was recorded. The temperature of the chamber was controlled by flowing cold nitrogen gas. All the samples were coated with a thin PVDF top-layer prior to the experiments, to ensure identical wetting properties (see Methods for details). The PVDF protective films exhibit very high optical transparency due to their sub-micron thickness. Moreover, they can act as single-layer anti-reflection coatings. This can be clarified by considering the relationship $n_{ar} = \sqrt{n_{air} n_{TiO_2}}$, where $n_{ar}$ is the refractive index of the single-layer anti-reflection coating, $n_{air} \approx 1$ is the refractive index of air and $n_{TiO_2} \approx 2.5$ is the refractive index of the enclosing TiO$_2$ layer of the metasurface in the visible wavelength range. This leads to the desired refractive index of the anti-reflection coating of $n_{ar} \approx 1.6$, which is close to the refractive index of PVDF, $n_{PVDF} \approx 1.35$.[37] Therefore, upon top-side illumination, we expect that the PVDF layer decreases the reflectivity on the top side of the surface, thus boosting the level of absorption and the plasmonic heating. This was also confirmed experimentally for a PVDF-coated partially-transparent metasurface with $L \approx 60$ nm, resulting in $\overline{\mathcal{A}} = 41\%$, vs. $\overline{\mathcal{A}} = 34\%$ for the same uncoated metasurface (see Supporting Information, section "Optical spectroscopy" and Figure S3 for the effect of the protection layer on the absorption, reflection and transmission of the metasurfaces). In our experiments, though, where bottom-side illumination is used, since the incident light does not pass through the PVDF layer before impinging on the metasurface, the levels of absorption and the plasmonic heating should not change considerably in the presence of the PVDF top-layer.

Figure 3b shows a plot of $\tau_{yx}$ vs. $t$ for the illuminated metasurface ($\overline{\mathcal{A}} = 37\%$, blue line) and control (black line) samples, for several de-icing cycles. The gray and blue shaded regions surrounding the blue (metasurface, 9 experiments) and black (control, 3 experiments) lines, respectively, are the minimum and maximum values of $\tau_{yx}$ observed during the experiments. Two regimes appear for the metasurface: almost constant $\tau_{yx}$ (prior to illumination) and sharply decreasing $\tau_{yx}$ (during illumination) until reaching the minimum measurable stress ($2.5 \pm 1.0$ kPa). For the control case, there is only one regime with constant $\tau_{yx}$. Figure 3c shows boxplots of de-icing times (i.e. time elapsed from $P > 0$ kW m$^{-2}$ until $\tau_{yx} \approx 0$), $t_d$, vs. $\overline{\mathcal{A}}$, at $T = -4$ °C. The mean de-icing times were $394 \pm 211$ s, $264 \pm 80$ s, $76 \pm 18$ s and $34 \pm 11$ s, for metasurfaces with $\overline{\mathcal{A}} = 28\%, 37\%, 63\%$ and $83\%$ respectively. From the graph, it is evident that there is an order of magnitude decrease in the de-icing time by increasing the amount of solar energy that is absorbed (metasurfaces with $\overline{\mathcal{A}} = 28\%$ vs. $\overline{\mathcal{A}} = 83\%$). For more information on the setup and calibration process of the force sensor, see Supporting Information, section 'Ice adhesion setup' and Figure S6. In the above, complete de-icing was achieved in all cases, which we attribute to the formation of an intervening melt layer at the surface. We ascribed the de-icing time, $t_d$, and the gradual reduction of $\tau_{yx}$ with time to the formation of a melt layer at the center of the ice–film contact area (warmest region) and subsequent radial outward propagation of the phase boundary towards the edge (coldest region). For considerations on the effect of viscous and capillary forces in re-sisting the ice–block motion, which we found to be insignificant relative to ice adhesion, see Supporting Information, section 'De-icing analysis'.

Figure 4a shows the cold chamber –integrated with the visible light illumination system– that was used to characterize the plasmonically enhanced light absorption heating in a partially transparent metasurface ($\overline{\mathcal{A}} = 37\%$, $L = 60$ nm) at sub-zero temperatures. For this part of the work, we chose to deposit the metasurface onto a thermally insulating substrate, poly(methyl methacrylate) (PMMA, $k_s = 0.2$ W m$^{-1}$ K$^{-1}$, $l = 1$ mm), in order to minimize thermal losses due to conduction. Both the control and metasurface were coated with a thin layer of PVDF (transparent) to ensure that the surface chemical composition is similar.

Next, we characterize the freezing behavior of a single supercooled water droplet on the illuminated control and metasurface. We ran the experiments by first placing the coated substrate on the sample holder. Then, we turned on the light source and focused it on the metasurface ($D \approx 6$ mm, $P \approx 2.4$ kW m$^{-2}$). Boiling liquid nitrogen was then flowed throughout the chamber to cool it down. We continuously measured the environmental gas ($T_1$) and surface ($T_2$) temperatures. To run the droplet freezing experiment, we first set $T_1 \approx -26$ °C. Then, a single water droplet, initially at room temperature, was deposited on the substrate at the center of the illuminated spot. Figure 4b-c shows representative side-view image sequences of a water droplet on a (b) control and (c) metasurface cooled down at a rate of $\approx 1$ °C min$^{-1}$, until spontaneous nucleation and freezing. This transition is characterized by a sudden change from a transparent to opaque droplet state (recalescent freezing,[40] see Supporting Information Video S1). Also indicated are the time, $t$, and $T_1$. Time-zero was considered as the time moment at which the droplet is in thermal equilibrium with the environment (to ensure that, we waited $\approx 5$ min after droplet placement and the light was switched on). The change in droplet volume during the experiments was relatively small and we estimated it to be $< 5\%$ h$^{-1}$. We define the environmental gas temperature just prior to freezing as $T_1^*$. The metasurface freezes at a much lower $T_1^*$ compared to the control ($t = 910$ s, $T_1^* = -48$ °C, vs. $t = 230$ s and $T_1^* = -33$ °C). Figure 4d shows a plot of $T_2$ vs. $T_1$ for the control (—) and metasurface (- - -). We see that the metasurface has a significantly higher temperature relative to the control case for a range of sub-zero temperatures (-53 to 30 °C). Figure 4e shows a plot (calibration curve) of $T_1^*$ vs. sample type (metasurface and control). For the control and metasurface, we measured $T_1^*$ to be $-34 \pm 2$ °C (16 experiments) and $-47 \pm 3$ °C (14 experiments), respectively. It is clear that there is a significant difference in $T_1^*$ for the two cases –which has equally significant implications in the freezing delay time (explored next)– that we can clearly attribute to the heating effect due to illumination.

To understand the significance of these results, we can use the classical nucleation theory.[17,41] We term the supercooled water droplet temperatures on an illuminated control and metasurface as $T_{d,0}$ and $T_d$, respectively. If we assume that $T_2 \approx T_{d,0}$ and $T_2 \approx T_d$ on the respective samples, and we set $T_1 \approx -34$ °C, then we have $T_{d,0} = -26$ °C and $T_d = -20$ °C (from the calibration curve). In the case of the control sample, $T_{d,0} = -26$ °C is the spontaneous nucleation temperature, $T_N$. The difference in droplet temperatures is then defined as $\Delta T = T_d - T_{d,0}$. Previously, it was shown that $\Delta T \propto \log_{10}(t_{av})$, where $t_{av}$ is the average time required for ice to nucleate in a supercooled droplet when the droplet is maintained at thermal equilibrium with its surroundings;[17] therefore, for a six degree temperature difference, one can expect a six orders of magnitude increase in $t_{av}$ for the metasurface relative to the control case, which is associated with a very pronounced freezing delay. See also Supporting Information, section 'Frosting characterization', Figure S7 and Figure S8 for frosting experiments in harsh environmental condi-tions: ambient humidity and high heat flux to the substrate.

Next, the defrosting potential of the partially transparent metasurfaces is investigated. Figure 5a–b shows an image sequence of a (a) frosted control sample and (b) metasurface ($\overline{\mathcal{A}} = 37\%$) samples (substrate: PMMA, $l = 1$ mm) in a cold dry environment ($T_1 = -16$ °C to $-15$ °C) that are illuminated with a halogen lamp ($P \approx 2.4 \text{ kW m}^{-2}$) for $0 \leq t \leq 600$ s. Figure 5a shows that the frost on the control sample is unaffected by illumination, while Figure 5b shows that the metasurface is completely defrosted at the illuminated area by $t = 140$ s (see Supporting Information Video S2 for a defrosting demonstration on a frosted control sample and metasurface). To eliminate the effects of surface composition on frost growth, both the control and metasurface were coated with a thin layer of PVDF. Frost was grown on both samples under identical environmental conditions and for the same duration, ensuring similar frost thicknesses. Due to the lower thermal conductivity and increased thickness of the PMMA substrate relative to the glass, we should expect that heat transfer into the sample holder should be minimized. Furthermore, we note that there is an insulating frost layer on top of the sample; therefore, one should expect a higher steady-state temperature increase in the illuminated metasurface, allowing defrosting to occur in-spite of the relatively cold surrounding environment.

**Conclusions**

In closing, we showed that with rationally designed hybrid metamaterial films that balance transparency and absorption, extreme icephobic surface performances can be achieved. Such films, here nanocomposites of gold and titanium dioxide, exhibit broadband visible light absorption, while being sub-wavelength thin, enabling localized ice melting at the film–ice interface. The plasmonically enhanced light absorption heating induced a temperature increase greater than 10 °C, compared to a control surface, for rapid de-icing within 30 s. Furthermore, we achieved a 6 °C decrease in the spontaneous nucleation temperature (resulting in ≈ 6 orders of magnitude increase in droplet freezing delay at -32 °C), and a defrosting time of ≈ 4 min, for a highly transparent metasurface ($\overline{\mathcal{A}} = 37\%$, $\overline{\mathcal{T}} = 36\%$). We presented a viable, passive, anti-icing and de-icing metamaterial platform harvesting the benefit of solar radiation, that can find a broad range of applications, especially where transparency is required, including water solar heating, automotive industry, residential and commercial buildings and construction or machinery infrastructure. We believe that although the present approach demonstrates both anti-icing and de-icing behavior while maintaining transparency, it could be further improved by incorporating other passive icephobicity designs based on surface nanoengineering.[16–24]

**Methods**

**Substrate preparation.** Double side polished, 4-in fused silica wafers ($l = 500$ μm) were sourced from UniversityWafer, Inc. A 5-μm protective photoresist layer was spin coated and developed on each wafer, which was subsequently cut into 18 mm by 18 mm square pieces, using an ADT ProVectus LA 7100 semi-automatic wafer dicer. The cut glass substrates were sonicated in acetone for 3 min, in order to remove the photoresist, followed by an equal-time sonication in isopropyl alcohol. Finally, they were dried in a nitrogen stream. PMMA substrates were prepared by manually cutting a PMMA sheet (Schlösser GmbH, $l = 1$ mm) into rectangular pieces (≈ 18 mm by 18 mm), removing the protective membrane and sonicating in water.

**Adhesion layer and thin film deposition.** A 2-nm chrome adhesion layer was deposited on the substrates, using an Evatec BAK501 LL thermal evaporator. The multilayer structure was then applied layer-by-layer *via* sputter deposition in argon atmosphere, by employing a Von Ardenne CS 320 C sputter tool. An RF field at a power of 600 W was used at the $TiO_2$ target, while a 50 W DC field was used in the case of the Au target. Deposition times were 43 s and 3 s, re-spectively, at a pressure of 6 μbar. A pre-sputtering time of 30 s was necessary for stabilizing the plasma and thus the deposition rate in the chamber. The first layer was TiO$_2$, followed by Au, and the alternation continued until the desired number of layers was reached. The deposition time for the TiO$_2$ top-layer was 72 s.

**Film characterization.** The samples were cleaved after scratching the glass substrate with a diamond tip on two opposite sides. Film thickness was extracted from cross-sectional images of the metasurface with $L = 270$ nm (44 layers), taken by a FIE Nova NanoSEM 450 scanning electron microscope, at an acceleration voltage of 2 kV. This approach also gave us a visualization of the cross-sectional particle distribution of the same metasurface. In the case of the top-view image of the Au nanoparticles (8-layer metasurface, $L = 45$ nm), acceleration voltage was 0.5 kV. For the particle size distribution (equivalent diameter) analysis of the top-view image, ImageJ and MATLAB software packages were employed. Light absorption measurements took place in two steps: the optical transmission, $\mathcal{T}$, and reflection, $\mathcal{R}$, spectra of the metasurface were individually recorded at the same spot on the sample, over the 400–800 nm wavelength range, by a UV–Visible spectrometer (Acton SP2500, Princeton Instruments), making the assumption of negligible light scattering. The absorption spectra were obtained by $\mathcal{A} = 1 - \mathcal{R} - \mathcal{T}$. Measurements from three different spots per sample were averaged to extract the absorption curves in Figure 1d and Figure S3a–d.

**Polymer protective coating and characterization.** A polyvinylidene fluoride (PVDF) protective layer was spin coated (Laurell WS-400B-6NPP/LITE) on top of the metasurfaces deposited on fused silica substrates, to minimize surface–ice interactions and provide mechanical durability. For this purpose, a PVDF solution (4 wt.%) in N,N-dimethylformamide (DMF) was prepared under rigorous stirring for 2.5 h. After cleaning the sample with acetone, under sonication, and isopropyl alcohol, the solution was spin coated (30 s at 3000 rpm) onto it, and then it was heated for 3 h at 200 °C, over the melting temperature of PVDF, to reduce surface roughness. The supplier of the PVDF (beads, $M_w \approx 180,000$) and DMF (anhydrous, 99.8%) was Sigma-Aldrich Co LLC. Advancing and receding water contact angle measurements were performed in a OCA 35 goniometer (DataPhysics), using the inflation/deflation technique (droplet volume of 8–10 μL), equal to $86.7 \pm 0.9\,°$ (advancing) and $73.0 \pm 1.3\,°$ (receding). The respective contact angles of a PVDF-coated only control substrate were $87.9 \pm 0.5°$ and $72.3 \pm 0.9\,°$. In the case of metasurfaces on PMMA substrates, a similar process to the ones on fused silica was followed, with the differences of only cleaning with sonication in water and heating up to 80 °C for 3 h (below the glass transition temperature of PMMA). A 4 wt.% PVDF solution in NMP (anhydrous, 99.5%, Sigma-Aldrich Co LLC) was used in this case.

**Simulation software.** MATLAB software suite was used to estimate the imaginary part of electric permittivity of our nanocomposite films. The heat transfer simulations were performed numerically in COMSOL Multiphysics Modeling Software.

**IR thermal response measurements.** The transient thermal response of the unprotected metasurfaces on fused silica substrate ($l = 500$ μm), as well as the one of an uncoated reference fused silica substrate, were measured by means of an infrared camera (FLIR SC7500, 1.5–5.1 um), equipped with a 50 mm F/2 lens, within 320 by 256 pixels (pixel pitch: 30 μm), at a framerate of 50 fps. A visible light illumination source (FLEXILUX 600 Longlife) consisting of a 50 W halogen lamp and a 5-mm diameter gooseneck fiber constituted the power source for the illumination of the samples. Light from the bottom side (substrate), was collimated and focused on the top-surface with two convex, 2-in lenses, using a Thorlabs monochrome CCD camera (DCC1545M-GL). The light spot had a diameter of $6.0 \pm 0.3$ mm, corresponding to a maximum power density of $2.4 \pm 0.2$ kW m$^{-2}$ (suns), measured with a Thorlabs S301C, 0.19–25 μm power meter. For calculating the emissivity of the samples, these were heated up on a hot plate to three discrete elevated temperatures, while the hot plate was kept at a low angle (less than 5°) with respect to the IR camera lens, in order to eliminate the Narcissus effect. A fast mechanical shutter (Melles Griot, 04 IMS 001) was used to cut the illumination on and off. The light was switched on at least 15 min prior to the experiments, for stabilization reasons. Five experiments were performed per sample, with a recording time of 200 s.

**De-icing time and ice adhesion measurements.** A home-built, temperature-controlled, zero humidity chamber was used for the purpose of the de-icing experiments under illumination, consisting of: a bronze cooling pipe, where cold nitrogen at -150 °C was supplied by a Kaltgas cryogenic cooling system; a piezo-actuated stage made by Smaract, consisting of three SLC-1730-S positioners; two 4-wire, class A, RTD temperature sensors (Pt-1000, class B, Sensirion); a humidity module (SHT30, Sensirion); a A201-1 piezoelectric force sensor (FlexiForce Quick-Start Board, Tekscan); a force transfer pin; and finally a 3-mm thick glass window that enabled the de-icing experiments due to illumination. The sensor values were recorded through a custom data acquisition box (Beckhoff). A hydrophobic polypropylene cylinder ($R = 1.5$ mm) was filled with fresh deionized water (EMD Millipore Direct-Q 3) and placed on the sample, which was mounted on the stage. The force transfer pin–sample distance was $\approx 1$ mm. A vacuum-insulated double shell minimized thermal losses and forced convection inside the chamber was enabled with a fan. Humidity levels were kept at zero throughout the experiments, *via* a cold nitrogen recirculation stream. The force vs. displacement data were then recorded every 50 ms. The maximum ice adhesion strength measurable with this setup for the given $R$ is $\approx 280$ kPa.

**Anti-icing and defrosting experiments.** The same chamber as in the de-icing experiments was used. A cold nitrogen recirculation stream maintained dry conditions. The exposure time of the camera was necessary to be readjusted, due to severe changes in the intensity of incident light, at the following time moments (defrosting experiments): (a) $t = 0$ s and (b) $t = 600$ s.

In the case, again, of the defrosting experiments (control: 3 experiments, metasurface: 3 experiments), frost was grown in ambient humidity conditions, by placing each sample on a cold block, at a temperature of -50 °C, for 45 min. Transfer of the frosted sample to the pre-cooled chamber was done in a fast manner to prevent melting of the formed layer.

**Anti-frosting experiments in ambient humidity conditions.** An in-house setup was prepared for the anti-frosting experiments, consisting of a xenon light source (300 W 6258 Xe lamp in a 87005 enclosure, Newport), two objectives (4x, 10x) to collimate and focus the light on the sample surface, which was vertically mounted on a peltier element (38.6 W, PE-127-14-25-S, Laird), a cooling system (SST-TD02-LITE, Silverstone) and a Thorlabs CCD camera (DCC1545M-GL). A PID peltier control circuit (TEC-1089-SV, TEC Engineering) was used to regulate temperature (measured with a PT-100 type RTD). The power density of light on the metasurface was $P \approx 1$ kW m$^{-2}$. The light was switched on for 30 min prior to the experiments for stabilization reasons. Recording framerate was 1 fps. The exposed area on the sample was dried with a nitrogen stream prior and after every frosting cycle.

## Supporting Information

The Supporting Information is available online on the ACS Publications website. The following sections are included: Modeling light absorption, nanoparticle size analysis, characterizing the absorption, transparency, and reflection of metasurfaces, thermography of the illuminated metasurfaces, heat transfer calculations, ice adhesion setup and de-icing analysis, and frosting characterization. Moreover, two videos are included, demonstrating the effect of illumination on droplet nucleation temperature, and defrosting of a partially transparent metasurface.

## Author Information

**Corresponding author**

E-mail : dpoulikakos@ethz.ch

E-mail : thomschu@ethz.ch

E-mail: eghlidim@ethz.ch

**Author Contributions Statement**

D.P. conceived the research idea, D.P., T.M.S., and H.E., designed research and provided scientific guidance in all aspects of the work. E.M., A.S., and C.H. conducted the experiments and analyzed the results. D.P., E.M., T.M.S., and H.E. wrote the paper draft and all authors participated in manuscript reading, correcting and commenting.
**Acknowledgements**
Partial support of the Swiss National Science Foundation under grant number 162565 and the European Research Council under Advanced Grant 669908 (INTICE) is acknowledged. E. M. thanks Reidt S. and Olziersky A. for their assistance in SEM image acquisition, Stutz R. for the sputtering deposition parameters, Drechsler U. for cleanroom introduction, Caimi D. for wafer dicing, Graeber G. for providing the blackbody sample, as well as Vidic J. and Feusi P. for technical support.

**Additional Information**

**Competing financial interests:** The authors declare no competing financial interests.

**References**

(1) Petrenko, V. F.; Higa, M.; Starostin, M.; Deresh, L. Pulse Electrothermal De-Icing. *Eng. Conf.* **2003**, *5*, 435–438.

(2) Gent, R. W.; Dart, N. P.; Cansdale, J. T. Aircraft Icing. *Philos. Trans. R. Soc. A Math.*

*Phys. Eng. Sci.* **2000**, *358*, 2873–2911.

(3) Farzaneh, M. *Atmospheric Icing of Power Networks*; Springer Netherlands, 2008.

(4) Rashid, T.; Khawaja, H. A.; Edvardsen, K. Review of Marine Icing and Anti-/de-Icing Systems. *J. Mar. Eng. Technol.* **2016**, *15*, 79–87.

(5) Norrman, J.; Eriksson, M.; Lindqvist, S. Relationships between Road Slipperiness, Traffic Accident Risk and Winter Road Maintenance Activity. *Clim. Res.* **2000**, *15*, 185–193.

(6) Tobiasson, W.; Buska, J.; Greatorex, A. Ventilating Attics to Minimize Icings at Eaves. *Energy Build.* **1994**, *21*, 229–234.

(7) Parent, O.; Ilinca, A. Anti-Icing and de-Icing Techniques for Wind Turbines: Critical Review. *Cold Reg. Sci. Technol.* **2011**, *65*, 88–96.

(8) Fillion, R. M.; Riahi, A. R.; Edrisy, A. A Review of Icing Prevention in Photovoltaic Devices by Surface Engineering. *Renew. Sustain. Energy Rev.* **2014**, *32*, 797–809.

(9) Farag, A.; Huang, L.-J. *CFD Analysis and Validation of Automotive Windshield De-Icing Simulation*; 2003.

(10) MarketsandMarkets Research Private. Aircraft De-Icing Market worth $1.30 Billion by 2020 https://www.marketsandmarkets.com/PressReleases/aircraft-de-icing.asp (accessed Aug 8, 2017).

(11) MarketsandMarkets Research Private. Ice Protection Systems Market worth 10.17 Billion USD by 2021 https://www.marketsandmarkets.com/PressReleases/ice-protection-system.asp (accessed Aug 8, 2017).

(12) Yang, P.; Sun, P.; Chai, Z.; Huang, L.; Cai, X.; Tan, S.; Song, J.; Mai, W. Large-Scale Fabrication of Pseudocapacitive Glass Windows That Combine Electrochromism and Energy Storage. *Angew. Chem., Int. Ed.* **2014**, *53*, 11935–11939.

(13) Martín-Palma, R. Silver-Based Low-Emissivity Coatings for Architectural Windows:

Optical and Structural Properties. *Sol. Energy Mater. Sol. Cells* **1998**, *53*, 55–66.

(14) Deng, X.; Schellenberger, F.; Papadopoulos, P.; Vollmer, D.; Butt, H. J. Liquid Drops Impacting Superamphiphobic Coatings. *Langmuir* **2013**, *29*, 7847–7856.

(15) Wong, T.-S.; Kang, S. H.; Tang, S. K. Y.; Smythe, E. J.; Hatton, B. D.; Grinthal, A.; Aizenberg, J. Bioinspired Self-Repairing Slippery Surfaces with Pressure-Stable Omniphobicity. *Nature* **2011**, *477*, 443–447.

(16) Schutzius, T. M.; Jung, S.; Maitra, T.; Eberle, P.; Antonini, C.; Stamatopoulos, C.; Poulikakos, D. Physics of Icing and Rational Design of Surfaces with Extraordinary Icephobicity. *Langmuir* **2015**, *31*, 4807–4821.

(17) Eberle, P.; Tiwari, M. K.; Maitra, T.; Poulikakos, D. Rational Nanostructuring of Surfaces for Extraordinary Icephobicity. *Nanoscale* **2014**, *6*, 4874–4881.

(18) Boreyko, J. B.; Srijanto, B. R.; Nguyen, T. D.; Vega, C.; Fuentes-Cabrera, M.; Collier, C. P. Dynamic Defrosting on Nanostructured Superhydrophobic Surfaces. *Langmuir* **2013**, *29*, 9516–9524.

(19) Davis, A.; Yeong, Y. H.; Steele, A.; Bayer, I. S.; Loth, E. Superhydrophobic Nanocomposite Surface Topography and Ice Adhesion. *ACS Appl. Mater. Interfaces* **2014**, *6*, 9272–9279.

(20) Meuler, A. J.; Smith, J. D.; Varanasi, K. K.; Mabry, J. M.; McKinley, G. H.; Cohen, R. E. Relationships between Water Wettability and Ice Adhesion. *ACS Appl. Mater. Interfaces* **2010**, *2*, 3100–3110.

(21) Chen, X.; Ma, R.; Zhou, H.; Zhou, X.; Che, L.; Yao, S.; Wang, Z. Activating the Microscale Edge Effect in a Hierarchical Surface for Frosting Suppression and Defrosting Promotion. *Sci. Rep.* **2013**, *3*, 2515.

(22) Kreder, M. J.; Alvarenga, J.; Kim, P.; Aizenberg, J. Design of Anti-Icing Surfaces:

Smooth, Textured or Slippery? *Nat. Rev. Mater.* **2016**, *1*, 15003.

(23) Kim, P.; Wong, T. S.; Alvarenga, J.; Kreder, M. J.; Adorno-Martinez, W. E.; Aizenberg, J. Liquid-Infused Nanostructured Surfaces with Extreme Anti-Ice and Anti-Frost Performance. *ACS Nano* **2012**, *6*, 6569–6577.

(24) Zhu, L.; Xue, J.; Wang, Y.; Chen, Q.; Ding, J.; Wang, Q. Ice-Phobic Coatings Based on Silicon-Oil-Infused Polydimethylsiloxane. *ACS Appl. Mater. Interfaces* **2013**, *5*, 4053–4062.

(25) He, X.; Liu, A.; Hu, X.; Song, M.; Duan, F.; Lan, Q.; Xiao, J.; Liu, J.; Zhang, M.; Chen, Y.; Zeng, Q. Temperature-Controlled Transparent-Film Heater Based on Silver Nanowire–PMMA Composite Film. *Nanotechnology* **2016**, *27*, 475709.

(26) Irajizad, P.; Hasnain, M.; Farokhnia, N.; Sajadi, S. M.; Ghasemi, H. Magnetic Slippery Extreme Icephobic Surfaces. *Nat. Commun.* **2016**, *7*, 13395.

(27) Zhou, L.; Tan, Y.; Wang, J.; Xu, W.; Yuan, Y.; Cai, W.; Zhu, S.; Zhu, J. 3D Self-Assembly of Aluminium Nanoparticles for Plasmon-Enhanced Solar Desalination. *Nat. Photonics* **2016**, *10*, 393–398.

(28) Clavero, C. Plasmon-Induced Hot-Electron Generation at Nanoparticle/Metal-Oxide Interfaces for Photovoltaic and Photocatalytic Devices. *Nat. Photonics* **2014**, *8*, 95–103.

(29) Boström, T.; Westin, G.; Wäckelgård, E. Optimization of a Solution-Chemically Derived Solar Absorbing Spectrally Selective Surface. *Sol. Energy Mater. Sol. Cells* **2007**, *91*, 38–43.

(30) Sarina, S.; Waclawik, E. R.; Zhu, H. Photocatalysis on Supported Gold and Silver Nanoparticles under Ultraviolet and Visible Light Irradiation. *Green Chem.* **2013**, *15*, 1814-1833.

(31) Li, X.; Xiao, D.; Zhang, Z. Landau Damping of Quantum Plasmons in Metal

Nanostructures. *New J. Phys.* **2013**, *15*, 023011.

(32) Brongersma, M. L.; Halas, N. J.; Nordlander, P. Plasmon-Induced Hot Carrier Science and Technology. *Nat. Nanotechnol.* **2015**, *10*, 25–34.

(33) Wu, K.; Chen, J.; McBride, J. R.; Lian, T. Efficient Hot-Electron Transfer by a Plasmon-Induced Interfacial Charge-Transfer Transition. *Science.* **2015**, *349*, 632–635.

(34) Qin, Z.; Wang, Y.; Randrianalisoa, J.; Raeesi, V.; Chan, W. C. W.; Lipinski, W.; Bischof, J. C. Quantitative Comparison of Photothermal Heat Generation between Gold Nanospheres and Nanorods. *Sci. Rep.* **2016**, *6*, 29836.

(35) Shalaev, V.; Cai, W. *Optical Metamaterials: Fundamentals and Applications*; Springer: New York, 2010.

(36) Hedayati, M. K.; Faupel, F.; Elbahri, M. Review of Plasmonic Nanocomposite Metamaterial Absorber. *Materials*. 2014, *7*, 1221–1248.

(37) Hedayati, M. K.; Javaherirahim, M.; Mozooni, B.; Abdelaziz, R.; Tavassolizadeh, A.; Chakravadhanula, V. S. K.; Zaporojtchenko, V.; Strunkus, T.; Faupel, F.; Elbahri, M. Design of a Perfect Black Absorber at Visible Frequencies Using Plasmonic Metamaterials. *Adv. Mater.* **2011**, *23*, 5410–5414.

(38) Incropera, F. P.; DeWitt, D. P.; Bergman, T. L.; Lavine, A. S. *Fundamentals of Heat and Mass Transfer*, 6th ed.; John Wiley & Sons, 2006.

(39) Perry, R. H.; Green, D. W.; Maloney, J. O. *Perry's Chemical Engineers' Handbook*, 7th ed.; McGraw-Hill: New York, 1997.

(40) Jung, S.; Tiwari, M. K.; Poulikakos, D. Frost Halos from Supercooled Water Droplets. *Proc. Natl. Acad. Sci.* **2012**, *109*, 16073–16078.

(41) Hobbs, P. V. *Ice Physics*; Clarendon Press: Oxford, 1974.

**Figures**

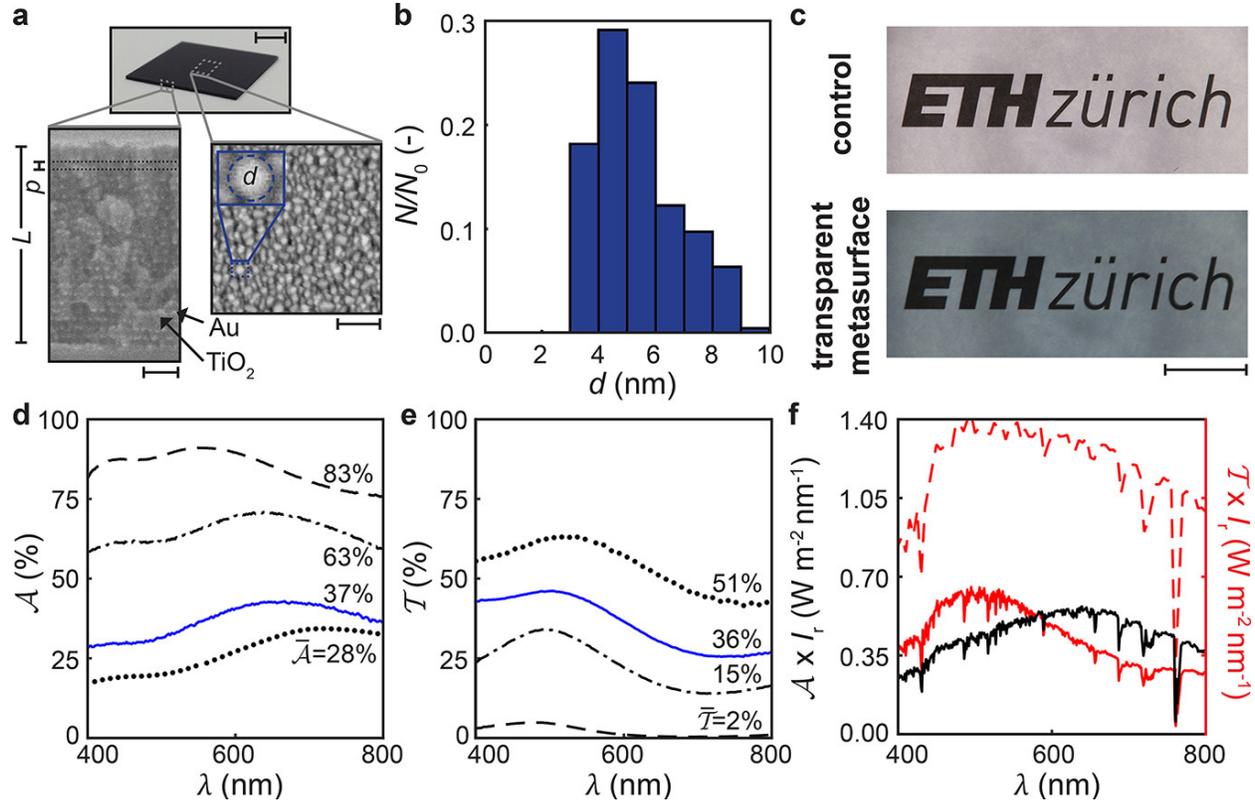

**Figure 1** Characterizing the topography and optical properties of the plasmonic metasurfaces. (**a**) Macroscopic (top-row) and microscopic (bottom-row) images of a metasurface; cross-sectional (bottom-left; $L = 270$ nm, $p = 11$ nm) and top-view (bottom-right; $L = 45$ nm) micrographs of the metasurface. Bright regions ($L = 45$ nm, obtained with backscattered and secondary electrons) correspond to gold nanoparticles. The volumetric concentration of gold is $\approx 40\%$. Sample surface is 18 mm by 18 mm. (**b**) Relative frequency of gold nanoparticles, $N/N_0$, vs. nanoparticle diameter, $d$ (sample properties: $L = 45$ nm, $p = 11$ nm). (**c**) Demonstration of the transparency of the metasurfaces (here, $L = 60$ nm), vs. a control sample, placed on a printed logo and under white backlight illumination. (**d**) Normalized absorption, $\mathcal{A}$, and (**e**) normalized transmission, $\mathcal{T}$, vs. wavelength of light (400–800 nm), $\lambda$, for films with varying $L$: 38 nm (···), 60 nm

(⎯), 95 nm (· – ·), and 270 nm (– – –). (**f**) Absorbed, $\mathcal{A} \cdot I_r$ (⎯), and transmitted, $\mathcal{T} \cdot I_r$ (⎯), sunlight (standard solar irradiance), vs. $\lambda$, for a metasurface ($L = 60$ nm) and reference sample ($\mathcal{A} = 0\%$, $\mathcal{T} = 100\%$, – – –). Scale bars: (**a**) top-row, 5 mm; bottom-left, 50 nm; bottom-right, 30 nm; (**c**) 2 cm.

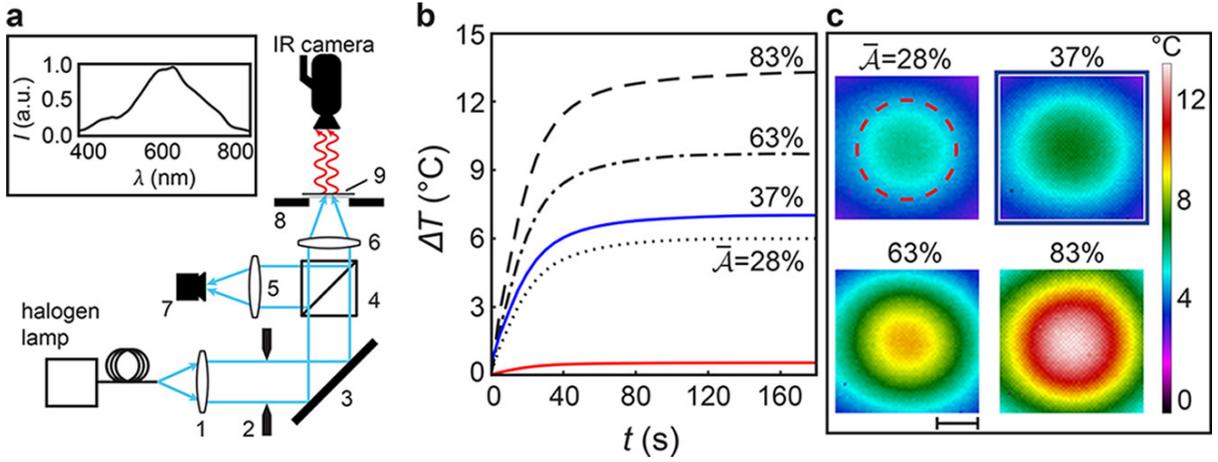

**Figure 2** Characterizing the heating behavior of the plasmonic metasurfaces (fused silica substrate) due to visible light exposure. (**a**) Schematic of the setup used to characterize the thermal response of the metasurfaces due to illumination: 1, collimating lens; 2, mechanical shutter; 3, silver mirror; 4, beam splitter; 5,6, focusing lenses; 7, CMOS camera; 8, sample holder; 9, sample. The spectrum of the broadband light source (⎯) is shown in the inset. (**b**) Temperature change, $\Delta T$, vs. time, $t$, for metasurfaces with varying values of $\overline{\mathcal{A}}$ (28% ···; 37% ⎯; 63% · – ·; 83% – – –) and a control substrate (⎯) after illumination ($P \approx 2.4$ kW m$^{-2}$); time-zero is defined as the moment that the mechanical shutter was opened. (**c**) Spatial distribution of $\Delta T$ at steady state ($t = 180$ s). The dashed circle represents the illuminated area, with a diameter of $D \approx 6$ mm. Scale bar: (**c**) 2.5 mm.

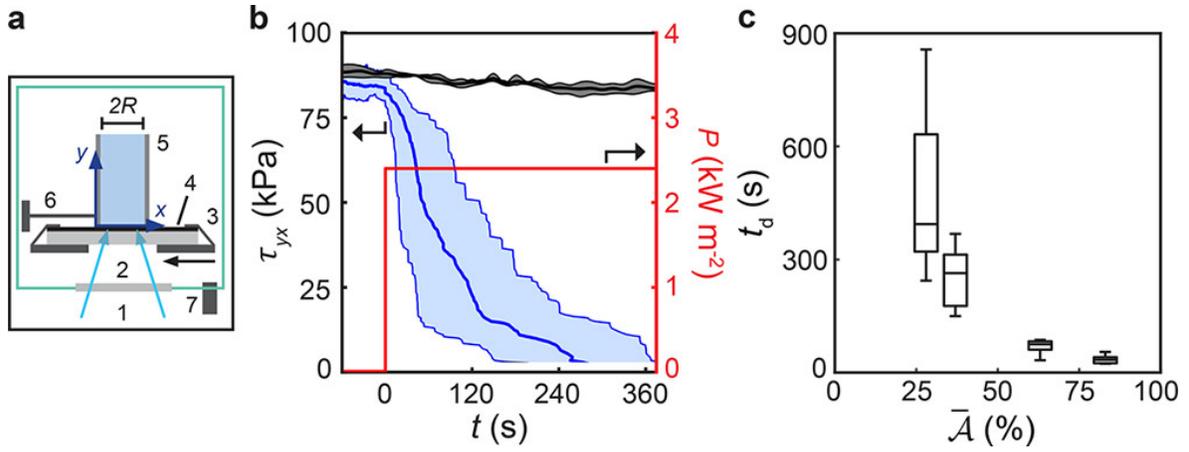

**Figure 3** Effect of visible light illumination on surface–ice adhesion. (**a**) Schematic of the setup used for measuring ice adhesion: 1, visible light illumination path, same as in the infrared temperature measurements; 2, glass window; 3, piezo-stage and sample holder; 4, sample; 5, non-wetting ice cylinder, with an inner radius of $R = 1.5$ mm; 6, piezoelectric force sensor (0–2 N, in-house calibration) and force transfer pin; 7, cold nitrogen vapor inlet. (**b**) De-icing curve (shear stress, $\tau_{yx}$, vs. time, $t$) of the sample with $\overline{\mathcal{A}} = 37\%$ (——). At $t = 0$ the sample was illuminated. The corresponding $\tau_{yx}$ vs. $t$ of a control sample (——) is also shown. The shaded areas show the minimum and maximum of the experimental measurements. (**c**) Boxplots of de-icing time, $t_d$ (time from maximum $\tau_{yx} \approx F_{max}/(\pi R^2)$ to noise level), vs. mean absorption, $\overline{\mathcal{A}}$, of the metasurfaces. The substrate was fused silica.

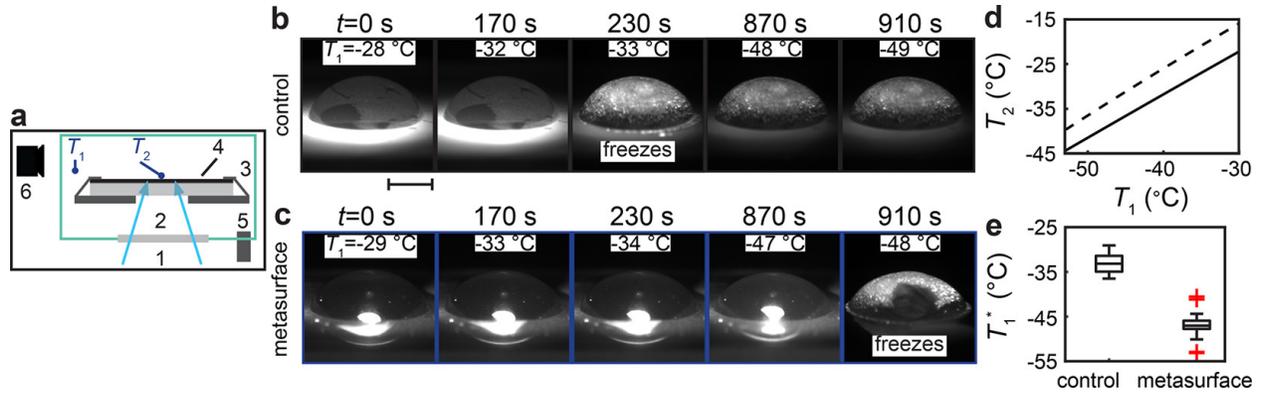

**Figure 4** Nucleation temperature of illuminated droplets. (**a**) Schematic of the environmental chamber used: 1, visible light illumination path; 2, glass window; 3, sample holder; 4, sample; 5, cold nitrogen vapor inlet; 6, CMOS camera. Two temperature sensors measured the gas ($T_1$) and sample ($T_2$) temperature. The illumination power density was $P \approx 2.4$ kW m$^{-2}$. Side-view image sequences of water droplets on a (**b**) control and (**c**) metasurface ($\overline{\mathcal{A}} = 37\%$) cooled down at a rate of $\approx 1$ °C min$^{-1}$; the final frames are when the droplets spontaneously nucleated and the second stage of freezing was progressing. The chamber gas temperature, $T_1$, is also shown. (**d**) Calibration curve: metasurface temperature, $T_2$, vs. gas temperature, $T_1$, in the case of the control (——) and metasurface (– – –). (**e**) Gas temperature at the moment of freezing, $T_1^*$, vs. sample type (control and metasurface). The substrate was PMMA. Scale bar: (**b**)–(**c**) 3 mm.

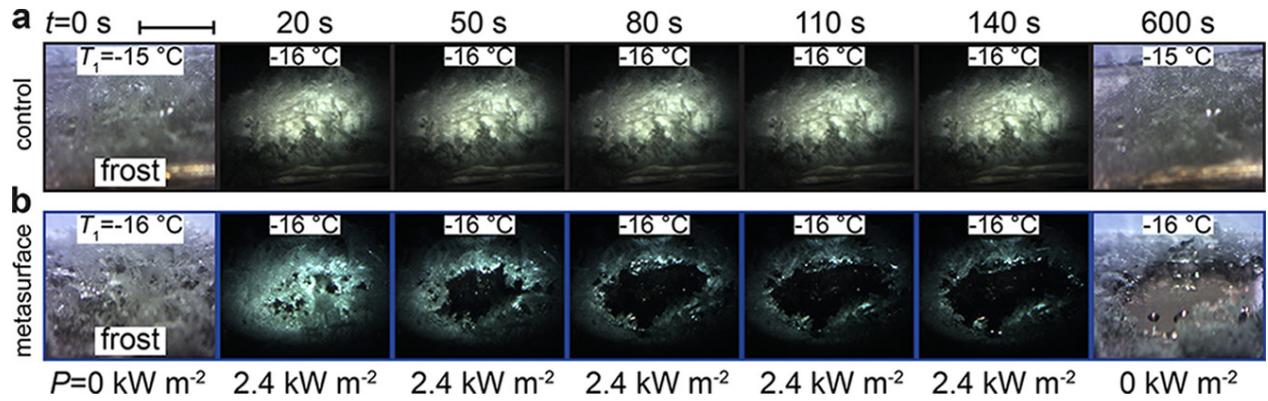

**Figure 5** Light-induced defrosting. Angled-view image sequences of frosted PVDF-coated (**a**) control and (**b**) metasurface ($\overline{\mathcal{A}} = 37\%$) samples that were illuminated with a halogen lamp ($P \approx 2.4$ kW m$^{-2}$) for $0 < t < 600$ s. At $t = 0$ s and $t = 600$ s, the samples were in ambient light conditions, revealing the frost before and after illumination. The chamber gas temperature, $T_1$, is also shown. The substrate was PMMA. Camera tilt angle was $\approx 25°$. Scale bar: (**a**)–(**b**) 4 mm.